\input{aipcheck}

\documentclass[
    ,final            % use final for the camera ready runs
  ]
  {aipproc}

\layoutstyle{6x9}

\newcommand{\be}{\begin{equation}}
\newcommand{\ee}{\end{equation}}

\newcommand{\bea}{\begin{eqnarray}}
\newcommand{\eea}{\end{eqnarray}}

\begin{document}

\title{Longitudinal and transverse meson correlators in the deconfined
phase from the lattice}



\author{Gert Aarts$^\ast$, \mbox{Chris Allton\footnote{Speaker}}\ }%$^\sharp$}
{address={Department of Physics, Swansea University, Swansea SA2 8PP,
United Kingdom}}

\author{Justin Foley}
{address={Department of Physics, Carnegie Mellon University,
Pittsburgh, PA 15213,USA}}

\author{Simon Hands}
{address={Department of Physics, Swansea University, Swansea SA2 8PP,
United Kingdom}}

\author{Seyong Kim}
{address={Department of Physics, Sejong University, Seoul 143-747,
Korea
%\\\vspace*{2mm}$\sharp$ Speaker
}}

\begin{abstract}
It has long been known that QCD undergoes a deconfining phase
transition at high temperature. One of the consequent features of this
new, quark-gluon phase is that hadrons become unbounded. In this talk
meson correlation functions at non-zero momentum are studied in
the deconfined phase using the Maximum Entropy Method.
\end{abstract}

\maketitle

%}}}

%{{{ Introduction

\section{Introduction}
\label{sec:intro}

\begin{figure}
\label{fig:qcd_phase}
  \includegraphics[height=.3\textheight]{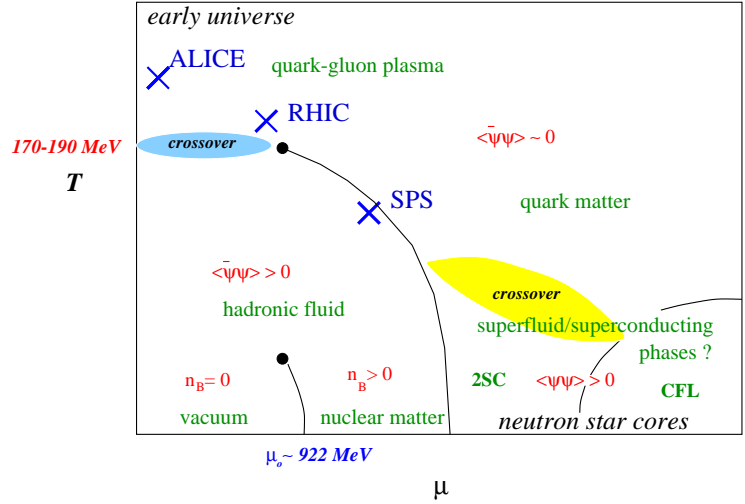}
  \caption{Illustrative phase diagram of QCD.}
\end{figure}

QCD is well-known to be a strongly interacting and confining theory
under ``normal'' conditions. However, it has long been theorised that
at large energy scales (in temperature, $T$, or baryonic chemical
potential, $\mu$) it undergoes a transition to a deconfined
(quark-gluon plasma) phase. This has been experimentally observed at
CERN \cite{SPS} and at the RHIC experiment at
Brookhaven \cite{RHIC}. The exact nature of this transition is
undergoing intense theoretical and experimental investigation.  In
fact, even the general features of the QCD phase diagram are still
being mapped out (see Fig.\ref{fig:qcd_phase}).  As an example of this
uncertainty, the Particle Data Book \cite{Amsler:2008zzb} does not
contain a single reference to the deconfined phase of QCD!  Properties
of QCD at $T=\mu=0$ are measured experimentally and calculated
theoretically often to accuracy at the percent level or below. The
same is certainly not true of the deconfined phase, where properties
typically have much larger ($\sim 20\%$) errors associated with them,
if they are known at all.

Physics at $T=\mu=0$ is characterised by quantities such as hadronic
masses and transition matrix elements, whereas the relevant quantities
in the deconfined phase correspond to those of plasma physics:
pressure, entropy, susceptibilities, and response functions. Of
particular interest in this work are transport coefficients.

From general arguments based on the asymptotically free nature of QCD,
the naive expectation is that quarks and gluons would be virtually
free in the deconfined phase. However, at RHIC, a (relatively)
strongly interacting phase was found with \lq\lq almost
instantaneous\rq\rq\ equilibration and a small ratio of viscosity,
$\eta$, to entropy density, $s$ (both signs of ``strong'' coupling).
These are characteristics of a so-called ``Perfect Fluid''.

There is some theoretical basis for such small values of $\eta/s$
in the plasma phase.  A calculation relying on the
correspondence between ${\cal N} = 4$ supersymmetric Yang-Mills theory
and superstring theory in $AdS_5 \times S^5$ space-time predicts a
lower bound for $\eta/s$
\cite{Policastro:2001yc,Kovtun:2004de},
\be
\label{eq:son}
\eta / s \ge \frac{1}{4\pi} \;\;\;\;\;\;\;\;\;
\mbox{for }\;\;N_c,\; g^2 N_c \rightarrow \infty.
\ee
While this theory is not QCD, it gives a scale for this ratio in a
strongly interacting theory \cite{Kovtun:2004de}.

Transport coefficients are essential properties of the plasma phase
and can be derived theoretically from current-current spectral
functions. We list some of them here.

\begin{description}

\item[Shear viscosity,] $\;\eta$,
which is obtained from off-diagonal energy-momentum correlators
in the zero energy limit \cite{Meyer:2007ic}.

\item[Bulk viscosity,] $\;\xi$,
which is obtained from diagonal energy-momentum correlators
in the zero energy limit \cite{Meyer:2007dy}.

\item[Electrical conductivity,] $\;\sigma$, and {\bf Diffusivity,} $\;D$,
which are obtained from the energy dependence of vector meson spectral
functions \cite{Aarts:2007wj}.
\end{description}

The aim of this talk is to make progress in the understanding of the
deconfining mechanism by analysing mesonic systems simulated by a
lattice calculation at $T \ne 0$ (but with $\mu = 0$). Naively, it is
expected that the two (valence) quarks in these mesons should become
unbound at the deconfining temperature, $T_c$. In practice this
appears true for light quark states only, and charmonium states appear
to remain bound until temperatures higher than
$T_c$ \cite{Aarts:2007pk, others}.  Specifically we aim to determine
the diffusivity through a lattice simulation of correlation functions
of vector currents at non-zero momentum. This work extends our
previous studies at zero momentum \cite{Aarts:2007wj} where we
calculated the electrical conductivity. A full version of this work is
in preparation \cite{future}.

%}}}

%{{{ Lattice Background

\section{Lattice Background}

The conventional approach to studying hadronic quantities with the
lattice technique is via the imaginary-time dependence of Euclidean
correlation functions of operators with well-defined quantum
numbers. In the confined phase, each bound state (in the tower of
states with those quantum numbers) contributes $\sim e^{-E_it}$ to the
correlation function, $G(t)$, where $E_i$ is the state's energy.  At
large times, $t$, the lowest state dominates, so fitting $G(t)$ to an
exponential form can trivially (in theory) determine $E_0$.

In the deconfined phase, the situation is more subtle. Unbound states
no longer contribute pure exponential terms to $G(t)$, and it is more
appropriate to introduce the spectral function,
$\rho(\omega, \vec{p})$,
\be
\label{eq:rho}
G(t,\vec{p}) = \int_0^\infty \rho(\omega,\vec{p}) \; K(t,\omega) \; 
\frac{d\omega}{2\pi},
\ee
where the (lattice) kernel is defined
\be
\label{eq:kernel}
K(t,\omega) =
\frac{\cosh[\omega(t-1/(2T))]}{\sinh[\omega/(2T)]},
\ee
and we have allowed for a momentum dependence in the correlation
function $G$ and therefore in $\rho$.
As usual, the temperature, $T$ is the inverse temporal length, $1/(aN_t)$.

In the confined case, each non-decaying state $i$ contributes a delta
function, $\delta(\omega-E_i)$, to $\rho(\omega)$. A decaying state
would have a spectral feature of finite width, and an unbound state
would correspond to a continuous spectrum. By studying the temperature
dependence of spectral functions the transition from the bound to
deconfined phases can be observed.

As well as containing information on the stability or otherwise of
hadronic states, $\rho(\omega,\vec{p})$ also can be used to extract
transport coefficients (as described in Sec.1) and hydrodynamic
structure.

However, despite the importance of the spectral function, and its
simple definition in terms of the correlation function in
eq.(\ref{eq:rho}), it is notoriously difficult to extract. This is
because it is an example of an ill-posed problem: there are (in
general) more $\omega$ data points in $\rho(\omega)$ than there are
$t$ data points in the correlation function $G(t)$. The method which
has now become fairly standard to overcome this problem is the {\em
Maximum Entropy Method} (MEM) which is based on Bayesian statistics
(for a review, see \cite{hatsuda}). MEM is a very standard technique
in fields which require image reconstruction/deconvolution, such as
astronomy, crystallography, and the analysis of atomic/molecular
spectra.

This work is an extension to our earlier work \cite{Aarts:2007wj,
Aarts:2006cq,Aarts:2006wt} to non-zero momentum. Specifically, we
apply MEM to calculate $\rho(\omega,\vec{p})$ for mesonic correlation
functions on lattice data with parameters summarised in
Table \ref{tb:params}.  We used the quenched approximation with the
standard Wilson gluonic and staggered fermionic actions, full details
are given in \cite{Aarts:2007wj}.  Twisted boundary
conditions \cite{Bedaque:2004kc,Flynn:2005in} were used to allow a
finer resolution in momentum space for $G(t,\vec{p})$ (and therefore
$\rho(\omega,\vec{p})$). In all 21 different momenta combinations were
studied (ranging up to $|\vec{p}| \sim 10/L$) of which 17 are
non-degenerate.

We follow our earlier work \cite{Aarts:2007wj} where a singularity in
$K(\omega,t)$ as $\omega\rightarrow 0$ was corrected by a simple redefinition
of $K \rightarrow \omega K / (2 T)$.  This allows us to obtain
reliable $\rho(\omega,\vec{p})$ estimates, even in the
$\omega \rightarrow 0$ limit.

%{{{ Table tb:params

\begin{table}
\begin{tabular}{lccc}
\hline
%&&&\\
                 &                   & {\bf Cold}               & {\bf Hot} \\
%&&&\\
\hline
%&&&\\
Spatial Volume   & $N_s^3 \times N_t$ & $48^3 \times 24$    & $64^3 \times 24$ \\
Lattice spacings & $a^{-1}$           & $\sim 4$ GeV             & $\sim 10$ GeV \\
$T$              & $1/(aN_t)$        & $T \sim 160 $MeV$ \sim 0.62
T_c$ & $T \sim 420 $MeV$ \sim 1.5 T_c$ \\
Statistics       & $N_{cfg}$          & $\sim 100$               & $\sim 100$ \\
%&&&\\
\hline
\end{tabular}
\caption{Lattice parameters used in the simulation.}
\protect\label{tb:params}
\end{table}

%}}}

%}}}

%{{{ Results

\section{Results}

\begin{figure}
\label{fig:stag}
  \includegraphics[height=.3\textheight]{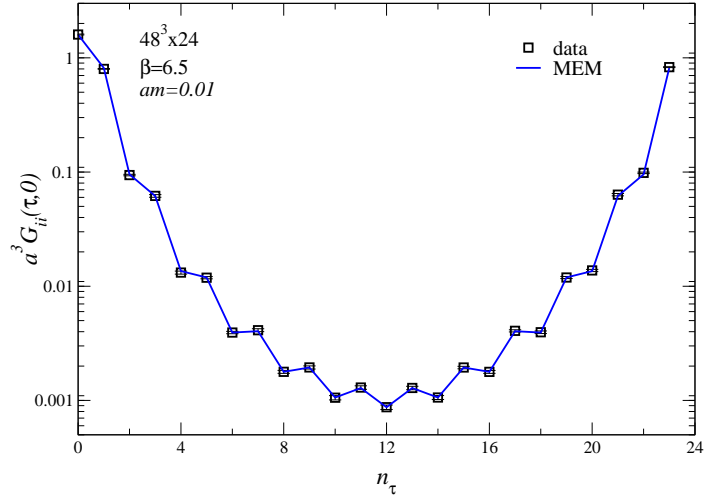}
  \caption{An example of the staggered correlation function showing
  the MEM analysis.}
\end{figure}

Our fermionic action uses staggered quarks so therefore the hadronic
correlators are a mixture of states with opposite parity, see
Fig.\ref{fig:stag}. These can be decomposed into the following spectral
representation,
\be
G(t,\vec{p}) = \int_0^\infty \frac{d\omega}{2\pi} K(t,\omega)
\left[ \rho(\omega,\vec{p}) - (-1)^{t} \; \overline{\rho}(\omega,\vec{p}) \right].
\ee
This means that to recover the physical spectral function, the even
and odd timeslices must be treated separately and then combined,
\be
\rho = \frac{1}{2}\left(\rho^{\mbox{even}} + \rho^{\mbox{odd}}\right).
\ee
Since there are only $N_t=24$ time slices in our lattices, the MEM
analysis of the even and odd timeslices includes six timeslices
(allowing for time reversal symmetry). This motivates the use of
anisotropic lattices in future studies \cite{Aarts:2007pk}.

\begin{figure}
\label{fig:momt}
  \includegraphics[height=.3\textheight]{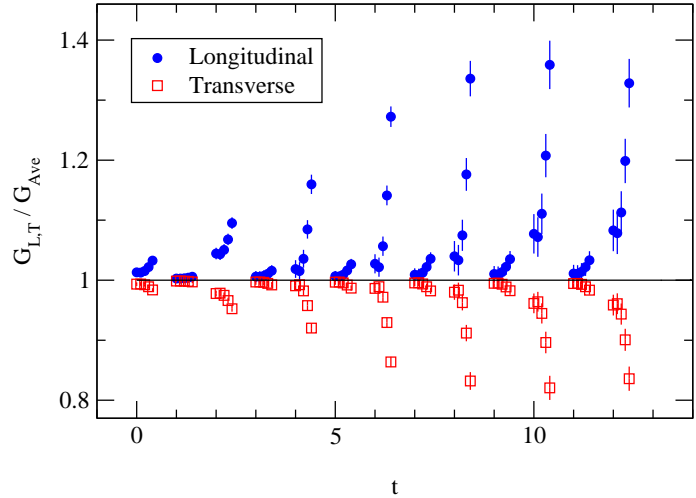}
  \caption{Longitudinal
  and transverse vector correlation functions normalised by the
  average correlation function,   $G_{\mbox{Ave}} = G_{\mbox{L}} + 2
  G_{\mbox{T}}$, for various momenta, $\vec{p}$, as a function of
  time.
  Data points for each momenta (for a given time) are offset
  horizontally for clarity; from left to right they are $|\vec{p}|L$ =
  0, 2, $\pi$, $2\pi$, $3\pi$.  }
\end{figure}

The electrical conductivity, defined
\be
\frac{\sigma}{T} =
\lim_{\omega\rightarrow 0} \frac{\rho_{ii}(\omega)}{6\omega T},
\ee
was found using this method \cite{Aarts:2007wj}, where $\rho_{ii}$ is
the spectral function for the spatial component of the vector
correlator.  The work presented here extends \cite{Aarts:2007wj} by
including non-zero momenta. In Fig.\ref{fig:momt}, we plot the
longitudinal and transverse vector correlation function versus time
for various momenta. As can be seen, there is a distinct difference
between the longitudinal and transverse correlators and a clear
systematic effect as the momenta increase.

\begin{figure}
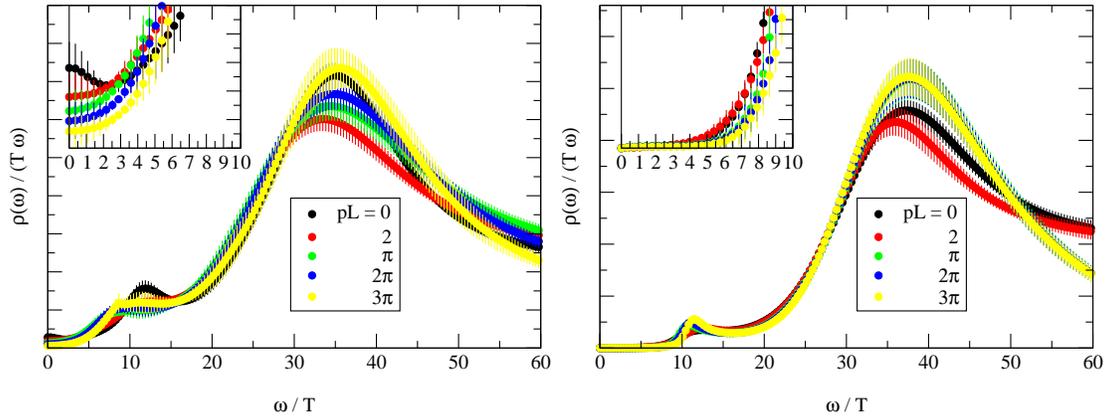

\label{fig:mom1}
  \includegraphics[width=.49\textwidth]{plot_long_m1_both.eps}
  \includegraphics[width=.49\textwidth]{plot_long_m2_both.eps}
  \caption{Longitudinal vector spectral function for various momenta
  in the hot case for $ma=0.01$ [left], and $ma=0.05$ [right].}
\end{figure}

The diffusivity, $D$, can be obtained, in principle, from the momentum
dependency of the longitudinal vector spectral function in the light
quark mass limit (see e.g. \cite{teaney}). In Fig. \ref{fig:mom1}, the
spectral functions for the longitudinal vector case are shown for both
the $ma=0.01$ and 0.05 quark masses. As can be seen, there is a clear
non-zero intercept in the case of the 0.01 mass which is absent in the
0.05 case. It is this non-zero intercept in the 0.01 case which led us
to determine the conductivity in
\cite{Aarts:2007wj}. Our future plans are to study this momentum dependency
with the aim of independently determining $D$ \cite{future}.

%}}}

%{{{ Acknowledgements

\begin{theacknowledgments}
CRA would like to thank the organiser of the workshop
``Achievements and New Directions in Subatomic Physics''.
S.K. was supported by the National Research Foundation of Korea grant
funded by the Korea government (MEST) No. 2009-0074027.
We also acknowledge the STFC grant ST/G000506/1.
\end{theacknowledgments}

%}}}

%{{{ References

%}}}

\end{document}